\documentclass[3p,11pt]{elsarticle}

\usepackage{graphicx,subfigure}
\usepackage{amssymb}
\usepackage{amsmath}
\usepackage{bm}
\usepackage{lineno} 
\usepackage{natbib}
\usepackage[squaren]{SIunits}
\usepackage{listings}
\journal{Comp. Phys. Comm.} 

\usepackage{color}
 
\definecolor{codegreen}{rgb}{0,0.6,0}
\definecolor{codegray}{rgb}{0.5,0.5,0.5}
\definecolor{codepurple}{rgb}{0.58,0,0.82}
\definecolor{backcolour}{rgb}{0.95,0.95,0.92}
 
\lstdefinestyle{mystyle}{
    backgroundcolor=\color{backcolour},   
    commentstyle=\color{codegreen},
    keywordstyle=\color{magenta},
    numberstyle=\tiny\color{codegray},
    stringstyle=\color{codepurple},
    basicstyle=\footnotesize\ttfamily,
    breakatwhitespace=false,         
    breaklines=true,                 
    captionpos=b,                    
    keepspaces=true,                 
    numbers=left,                    
    numbersep=5pt,                  
    showspaces=false,                
    showstringspaces=false,
    showtabs=false,                  
    tabsize=2,
}

\begin{document}

\begin{frontmatter}

\title{Controlling the Numerical Cerenkov Instability in PIC simulations using a customized finite difference Maxwell solver and a local FFT based current correction}

\author[THUACC]{Fei Li} 
\author[UCLAEE]{Peicheng Yu}
\ead{tpc02@ucla.edu}
\author[UCLAEE,UCLAPH]{Xinlu Xu}
\author[SLAC]{Frederico Fiuza} 
\author[UCLAPH]{Viktor K. Decyk}
\author[UCLAPH]{Thamine Dalichaouch}
\author[UCLAPH]{Asher Davidson}
\author[UCLAPH]{Adam Tableman}
\author[UCLAEE,UCLAPH]{Weiming An}
\author[UCLAPH]{Frank S. Tsung}
\author[IST,ISCTE]{Ricardo A. Fonseca}
\author[THUACC]{Wei Lu}
\author[UCLAEE,UCLAPH]{Warren B. Mori}

\address[THUACC]{Department of Engineering Physics, Tsinghua University, Beijing 100084, China}
\address[UCLAEE]{Department of Electrical Engineering, University of California Los Angeles, Los Angeles, CA 90095, USA}
\address[UCLAPH]{Department of Physics and Astronomy, University of California Los Angeles, Los Angeles, CA 90095, USA}
\address[SLAC]{SLAC National Accelerator Laboratory, Menlo Park, CA 94025}
\address[IST]{GOLP/Instituto de Plasma e Fus\~ao Nuclear, Instituto Superior T\'ecnico, Universidade de Lisboa, Lisbon, Portugal}
\address[ISCTE]{ISCTE - Instituto Universit\'ario de Lisboa, 1649--026, Lisbon, Portugal}

\begin{abstract}
In this paper we present a customized finite-difference-time-domain (FDTD)  Maxwell solver for the particle-in-cell (PIC) algorithm. The solver is customized  to effectively eliminate the numerical Cerenkov instability (NCI) which arises when a plasma (neutral or non-neutral)  relativistically drifts on a grid when using the PIC algorithm. We control the EM dispersion curve in the direction of the plasma drift of a FDTD Maxwell solver by using a customized higher order finite difference operator for the spatial derivative along the direction of the drift ($\hat 1$ direction). We show that this eliminates the main NCI modes with moderate $\vert k_1 \vert$, while keeps additional main NCI modes well outside the range of physical interest with higher $\vert k_1 \vert$. These main NCI modes can be easily filtered out along with first spatial aliasing NCI modes which are also at the edge of the fundamental Brillouin zone. The customized solver has the possible advantage of improved parallel scalability because it can be easily partitioned along $\hat 1$ which typically has many more cells than other directions for the problems of interest. We show that FFTs can be performed locally to current on each partition to filter out the main and first spatial aliasing NCI modes,  and to correct the current so that it satisfies the continuity equation for the customized spatial derivative. This ensures that Gauss' Law is satisfied. We present simulation examples of one relativistically drifting plasmas, of two  colliding relativistically drifting plasmas, and of nonlinear laser wakefield acceleration (LWFA)  in a Lorentz boosted frame that show no evidence of the NCI can be observed when using this customized Maxwell solver together with its NCI elimination scheme. 
\end{abstract}


\end{frontmatter}


\section{Introduction}
\label{sect:intro}

When modeling physics problems involving relativistically drifting plasmas or particle beams using the electromagnetic (EM) particle-in-cell (PIC) algorithm, a violent numerical instability, called the numerical Cerenkov instability (NCI) \cite{Godfrey1974,Godfrey1975,YuAAC12,GodfreyJCP2013,XuCPC13}, arises due to the unphysical coupling of the electromagnetic (EM) modes and Langmuir modes (both main and aliasing) of the drifting plasma in the numerical system \cite{YuCPC15}. Such scenarios arise in laser wakefield acceleration \cite{LWFA} simulations in the Lorentz boosted frame \cite{MoriProposal,Vay2007PRL} in which a laser pulse is colliding head-on with a relativistically drifting plasma (henceforth moving in the $\hat 1$ direction), and in relativistic collisionless shock simulations in which two counter-propagating relativistically drifting plasmas collide head-on with each other \cite{shock1,shock2,shock3,shock4}.  The NCI creates unphysical energy exchange between the kinetic energy of the drifting plasma, and EM field fluctuations. The growth of the fastest growing modes drive fluctuations that can quickly dwarf the physical processes being studied, while slower growing modes can effect the physics in subtle ways. Thus, it is essential to eliminate all NCI modes for the accurate modeling of physics problems.

Recently, several studies have focused on the understanding and elimination of this instability \cite{YuAAC12,GodfreyJCP2013,XuCPC13,YuCPC15,GodfreyPS,GodfreyFDTD,YuCPC152,GodfreyGPS}. This includes the use of a multi-dimensional spectral solver \cite{YuCPC15,YuJCP14, lin, dawsonrmp}, or a hybrid Yee-FFT solver that uses a FFT in the direction of the drifting plasma \cite{YuCPC152}. Applying these solvers together with current correction and filtering strategies, one can systematically eliminate the NCI modes. The key idea is that by changing the EM dispersion relation of the solver it is possible to eliminate NCI modes at moderate  $\vert \vec k \vert$ leaving all the remaining NCI modes of interest at high $\vert \vec k \vert$ in the fundamental Brillouin zone. These high $\vert \vec k \vert$ modes can be filtered without altering the physics of interest. However, each of these solvers is FFT-based, either in all directions for the pure spectral solver, or only in the drifting direction of the plasma. The use of FFTs in the Maxwell solver can limit the parallel scalability of the algorithm when there are many more cells along a given  direction. Typically, in parallel PIC codes based on FFT solvers, one uses a domain decomposition one order lower than the physical dimension because there are currently no effective parallel 1D FFTs. In the hybrid Yee-FFT solver, there is only a 1D FFT, so the problem can only be decomposed in the two transverse directions. 

In this paper, we take advantage of the previous progress described in \cite{YuCPC15,YuCPC152}, and develop a method to design a finite-difference-time-domain (FDTD) solver that has similar (yet different) NCI properties to the FFT-based solver described in \cite{YuCPC15,YuCPC152}. Although it was based on the use of FFTs, when examined more carefully the previous progress showed that the key to essentially eliminate the NCI is to first  isolate the range of unstable $\vec k$'s for what we refer to as the main NCI mode. This is accomplished by using a solver that has sufficiently small numerical errors in the spatial derivatives (and thus small dispersion errors for light waves) for moderate $\vert k_1 \vert$.  Even with perfect dispersion for light waves in vacuum, there will always be an intersection from the first spatial aliased beam modes at high $\vert k_1 \vert$ that needs to be filtered out, and coupling between the EM mode and the main Langmuir mode at moderate $\vert k_1 \vert$. We note that for this reason the use of the PSATD solver described in Ref. \cite{GodfreyPS} does not appear to have advantages with respect to eliminating the NCI.  As we have recently shown \cite{YuCPC15}, when  the main mode is isolated to a small range of $\vec k$s then a small modification to the dispersion for the range of unstable modes can remove the coupling between the EM (purely transverse in the lab frame)  modes and Langmuir (purely longitudinal in the lab frame) modes. 

Recognizing how the NCI is being controlled and eliminated by the use of an FFT along the plasma drifting direction ($\hat 1$ direction) leads us to consider the possibility to design a customized and higher order finite difference operator for the spatial derivatives that provides sufficiently accurate dispersion for moderate  $\vert \vec k \vert$. This finite difference operator for the spatial derivative can be implemented into a FDTD solver which is purely local and should thus scale well on massively parallel computers using domain decomposition. While this new solver can eliminate unstable modes at moderate $\vert k_1 \vert$, it cannot eliminate modes at high $\vert k_1 \vert$ near the edge of the first Brillouin zone. In addition, such a solver will not conserve charge i.e., Gauss's law will not be satisfied. Nonetheless, both of these issues can be resolved by performing local FFTs  for the current which do not use any global communication.

Similar to the hybrid Yee-FFT solver, to ensure the Gauss' Law is satisfied for the customized solver, we correct the component of the current in the $\hat 1$ direction in $k_1$ space. This is done locally on each parallel partition along $\hat 1$ and because the current is already in  $k_1$ space, we can also use a low pass filter and eliminate the unstable high $k_1$  NCI modes. This filter can also be included into the current correction. We will show that overall this method is effective at eliminating the NCI while allowing good parallel scalability when domain decomposition is required along $\hat 1$.

We note that in Ref. \cite{VayJCP2013} a PIC algorithm based on using a ``local'' FFT Maxwell solver was proposed. Their work was motivated for maintaining high parallel efficiency and was not focused on eliminating the NCI. They did show results from LWFA simulations in a boosted frame, however, no analysis for the NCI for FFT based algorithms was presented. We note that there are distinct differences between their approaches to ours. In our case, FFTs are performed only on current. This is done to ensure that the continuity equation is satisfied, and also to filter the current for NCI elimination. Because the current from a single particle is not global this can lead to a current that satisfies the continuity equation at every location and it can eliminate the NCI. The EM fields remain in real space and are advanced using Faraday's and Ampere's law. So long as the solution for the current satisfies the continuity equation locally for the finite difference operators used in Ampere's Law, then Gauss's law will be maintained. If the longitudinal components of the fields are also solved using local FFTs in each partition (as is proposed in Ref. \cite{VayJCP2013}), there will be errors in the longitudinal components of $\vec E$ and $\vec B$ due to the enforcement of periodicity.

The remainder of this paper is organized as follows. In section \ref{sect:solver} we first present a method to construct a customized FDTD Maxwell solver that has preferable NCI properties. The corresponding current correction and filtering strategies are discussed in section \ref{sect:current}. We show that the use of  local FFTs can provide a current that satisfies the continuity equation for a customized solver. We then present sample simulations in section \ref{sect:simulation} showing that good accuracy and scalability can be obtained. Finally, in section \ref{sect:summary} a summary is given.

\section{Customized Maxwell solver}
\label{sect:solver}
The Numerical Cerenkov Instability (NCI) occurs when a plasma drifts relativistically on a grid in a PIC code due to the unphysical coupling between the Langmuir modes (both main and aliasing) and electromagnetic (EM) modes \cite{YuCPC15}. Categorizing the NCI modes with their temporal aliasing mode number $\mu$, and spatial aliasing mode number $\nu_1$, it is found that usually the most violently growing NCI modes are those at $(\mu,\nu_1)=(0,\pm 1)$ (we call them first spatial aliasing NCI modes), and $(\mu,\nu_1)=(0,0)$ (main NCI modes) \cite{YuCPC15}. The first spatial NCI modes usually reside near the edges of the fundamental Brillouin zone, making them relatively easy to eliminate with a sharp low-pass filter to the current. On the other hand, the main NCI modes are usually located within the inner half of the $\vec k$ modes, where  modes of physical interest are located. It was shown for typical FDTD solvers that these modes were contained in a broad spectrum such that they cannot be eliminated through a low pass or mask filter. On the other hand, as shown in \cite{YuCPC15,YuCPC152}, for FFT-based solvers (and cell sizes $\Delta x_1 \le \Delta x_2$), the main NCI modes are very localized in $\vec k$ space and they move to large $\vert \vec k \vert$ as the time step is reduced. 

Therefore, as discussed in \cite{YuCPC15,YuCPC152}, to eliminate the NCI modes in FFT-based solvers, the first step is to find a reduced time step which moves the main NCI modes away from the physical modes. After these modes are far enough from the physical modes, one can then apply a highly localized modulation to the EM dispersion relation in $\vec k$ space  where the $(\mu,\nu_1)=(0,0)$ modes reside in order to completely eliminate them. For a multi-dimensional spectral solver, the modification of the EM dispersion is accomplished by directly modifying the finite difference operator $[\vec k]$ in that localized area in $\vec k$ space \cite{YuCPC15}. For the hybrid Yee-FFT solver, only the operator $[k_1]$ is modified in the $k_1$ range where the main NCI modes are located \cite{YuCPC152}. The modification of the operator usually creates a bump in the dispersion curve in the range where the main NCI modes are located, which removes the coupling between the EM modes and the Langmuir modes in that area, thereby eliminating the main NCI modes in that range completely.

When solving the Maxwell's equation using the FFT-based solvers, the differential operators in $\vec k$ space are explicitly used in the equations, therefore it is straightforward to modify the operators in $\vec k$ space. However, as mentioned in introduction, the use of a 1D FFT when there are many cells along that direction affects the scalability of the solver (or a multi-dimensional FFT solver when there are ``many'' more cells along one direction). Therefore, a question that naturally follows is whether it is possible to design an FDTD solver to imitate the characteristics of the EM dispersion curves of a FFT-based solver that make it possible to effectively eliminate the NCI. In the following sections, we describe a ``recipe'' for designing a finite difference derivative that when written in $\vec k$ space leads to the proper characteristics for the EM dispersion. 

In \cite{YuCPC152} it is found that by replacing the finite difference spatial derivative in the direction of the plasma drift from a stencil that is second order accurate in cell size with a spectral solver  (which is greater than $N$-th order accurate), one can restrict the $(\mu,\nu_1)=(0,0)$ NCI modes to a highly localized area in the fundamental Brillouin zone \cite{YuCPC152}. Meanwhile, the spatial derivatives in the other direction(s) can be kept as second order accurate. Therefore, when we design an FDTD solver for the purpose of NCI elimination, it is natural to start with a ``hybrid'' FDTD solver that resembles the hybrid Yee-FFT solver. We use a higher order FDTD finite difference stencil \cite{Khan99, Khan03} in the direction of the drift  while keeping them second order accurate in the other direction(s). Examination of the NCI growth rate where $[k_1]$ is replaced with the expression for a higher order stencil reveals that indeed the NCI is localized. In addition, we find that new  $(\mu,\nu_1)=(0,0)$ modes arise at large $k_1$ where the  EM dispersion curve rolls over, i.e., the phase velocity drops.  We then show how to modify the expression $[k_1]$ for the higher order finite difference operator such that the EM dispersion curve has a slight bump at moderate $\vert k_1\vert$ in order to precisely avoid the coupling  between the EM modes and main Langmuir modes for the main, $(\mu,\nu_1)=(0,0)$, NCI modes of moderate $\vert k_1\vert$. To accomplish that, we expand the number of terms in the stencil [see Eq. (\ref{eq:cpl})] to add extra degrees of freedom which can create the bump in the $k_1$ range where the main NCI modes reside, as we will explain in the following sections. In addition the new $(\mu,\nu_1)=(0,0)$ modes at high $k_1$ can be filtered out since they are outside the range of physically relevant modes.

\subsection{NCI for high order finite difference solvers}
\label{subsect:NCI_high-order}
Without loss of generality, we describe the method outlined above in the 2D Cartesian geometry. In a Maxwell solver, the electromagnetic fields $\vec{E}$ and $\vec{B}$ are advanced by solving Faraday's law and Ampere's law,
\begin{eqnarray}
	\label{eq:maxwell_eq1}
	\vec{B}^{n+\frac{1}{2}} &=& \vec{B}^{n-\frac{1}{2}} -c\Delta t \nabla^+_p \times \vec{E}^n\\
	\label{eq:maxwell_eq2}
	\vec{E}^{n+1} &=& \vec{E}^n +c\Delta t \nabla_p^- \times \vec{B}^{n+\frac{1}{2}} - 4\pi\Delta t \vec{J}^{n+\frac{1}{2}}
\end{eqnarray}
where the EM fields $\vec{E}$ and $\vec{B}$, and current $\vec{J}$ are defined on the staggered Yee grid \cite{Yee},  and $\nabla_p^\pm=\left( \partial_{p,x_1}^\pm, \partial_{2,x_2}^\pm \right)$ is the discrete finite difference operator for the staggered scheme. Note according to \cite{YuCPC152} the NCI can be eliminated if the operator $[k_1]=k_1$ is used along the plasma drifting direction. We now show that for a FDTD solver, a similar dispersion curve can be obtained by using high order finite difference operator in this direction. We apply a $p$-th order operator in the $\hat 1$ direction and a standard second order Yee solver in the $\hat 2$ direction. The $p$-th order operator is defined as
\begin{eqnarray}
\label{eq:cpl}
	\partial_{p,x_1}^+f_{i_1,i_2} &=& \frac{1}{\Delta x_1}\sum_{l=1}^{p/2} C_l^p \left(f_{i_1+l,i_2}-f_{i_1-l+1,i_2}\right) \nonumber\\
	\partial_{p,x_1}^-f_{i_1,i_2} &=& \frac{1}{\Delta x_1}\sum_{l=1}^{p/2} C_l^p \left(f_{i_1+l-1,i_2}-f_{i_1-l,i_2}\right)
\end{eqnarray}
where $f$ is an arbitrary quantity and the coefficients of the finite difference operator $C_l^p$ can be expressed as \cite{Khan99, Khan03}:
\begin{equation}
	C_l^p = \frac{(-1)^{l+1}16^{1-\frac{p}{2}}(p-1)!^2}{(2l-1)^2(\frac{p}{2}+l-1)!(\frac{p}{2}-l)!(\frac{p}{2}-1)!^2}
\end{equation}
If we perform a Fourier transform to Eqs. (\ref{eq:maxwell_eq1}) and (\ref{eq:maxwell_eq2}) in both time and space, Maxwell's equations become
\begin{eqnarray}
[\omega]\vec{B} &=& -[\vec{k}] \times \vec{E}\\
\label{eq:maxwell_wk}
\ [\omega]\vec{E} &=& [\vec{k}] \times \vec{B} + 4\pi i\vec{J}
\end{eqnarray}
where the operators in operators in frequency and wavenumber space are
\begin{equation}
\label{eq:maxwell_wk1}
\begin{aligned}	
	\ [\omega]&=\frac{\sin(\omega\Delta t/2)}{\Delta t/2} \\
	[\vec{k}]&=\left( \sum_{l=1}^{p/2} C_l^p \frac{\sin[(2l-1)k_1\Delta x_1/2]}{\Delta x_1/2}, \frac{\sin(k_2\Delta x_2/2)}{\Delta x_2/2} \right)
\end{aligned}
\end{equation}
where $\omega$ and $k_{1,2}$ are the frequency and wave numbers, and $\Delta t$ and $\Delta x_{1,2}$ are the time step and grid sizes of the PIC system, respectively. Note that when the current vanishes, $\vec{J}=\vec{0}$,  in Eq (\ref{eq:maxwell_wk}), we obtain the numerical dispersion relation for EM waves in vacuum, i.e., 
\begin{equation}
	[\omega]^2 = c^2\left([k_1]^2_p+[k_2]^2_2\right)
\end{equation}
where $[k_1]_p$ and $[k_2]_2$ are the components of $[\vec{k}]$, and the order of accuracy is denoted by the subscripts outside the square brackets. 

\begin{figure}[hbtp]
\centering
\includegraphics[width=0.7\textwidth]{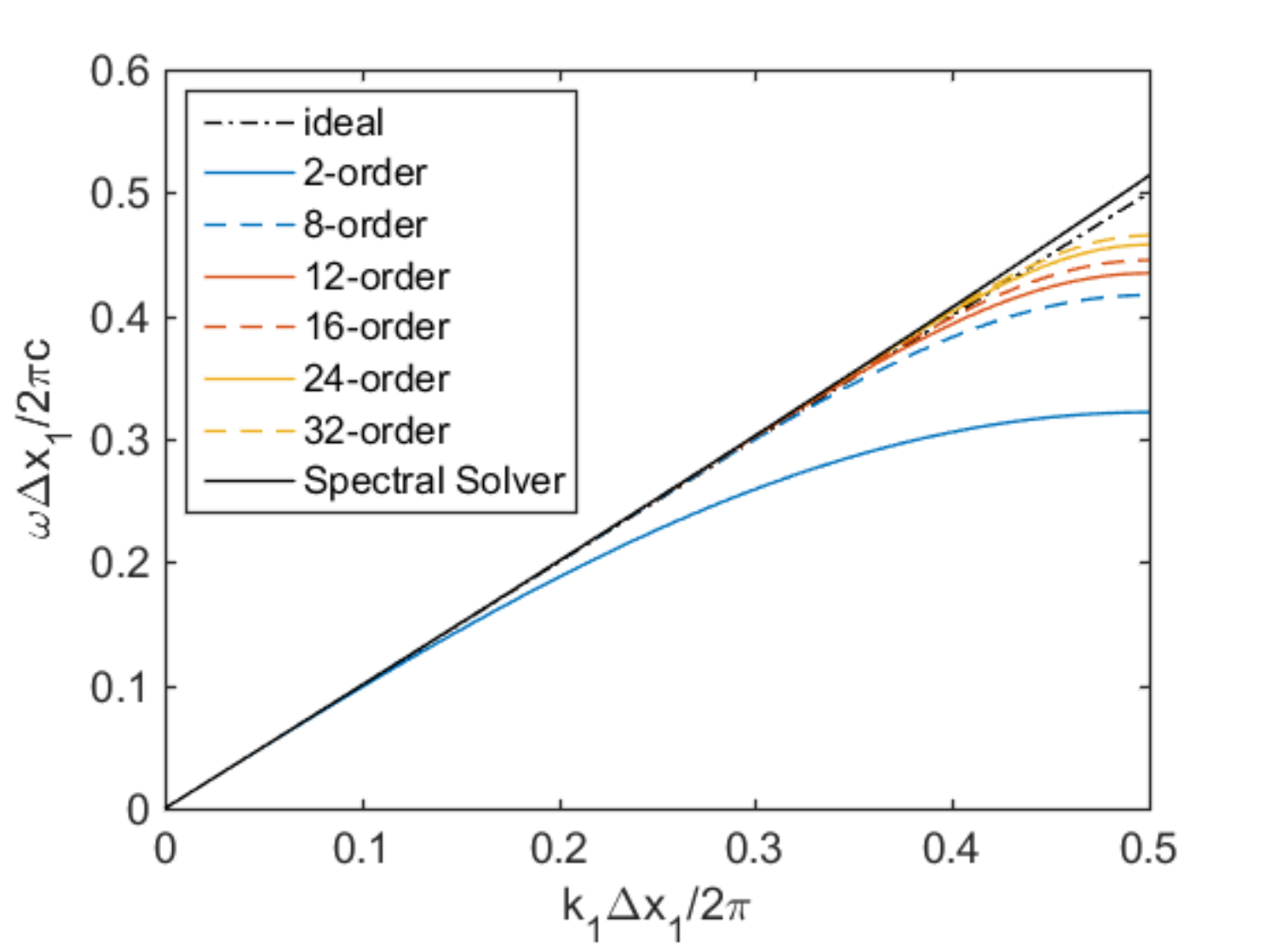}
\caption{1D numerical dispersion relations of different finite difference solver and spectral solver. $\Delta x_1=0.2k_0^{-1}$ and $\Delta t=0.05\omega_0^{-1}$ are used to generate this plot.}
\label{fig:1d_num_disp}
\end{figure}

\begin{figure}[hbtp]
\centering
\includegraphics[width=0.7\textwidth]{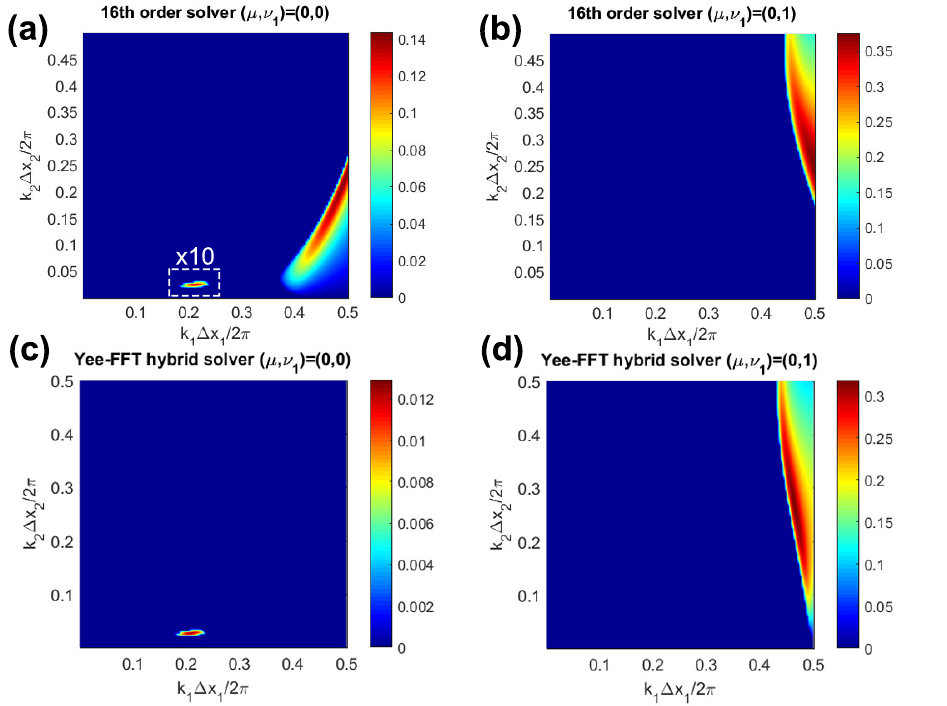}
\caption{The NCI patterns of 16th order solver and hybrid Yee-FFT solver. (a) and (b) show the $(\mu,\nu_1)=(0,0)$ and $(\mu,\nu_1)=(0,1)$ NCI mode of the high order (16th order) solver. The values in the dashed line box in (a) is multiplied by 10 for better visualization. (c) and (d) show the $(\mu,\nu_1)=(0,0)$ and $(\mu,\nu_1)=(0,1)$ NCI mode of the hybrid solver. $\Delta x_{1,2}=0.2$, $\Delta t=0.05$ and $n_p=30n_0$ are used to generate the plots.}
\label{fig:nci_16ord_hybrid}
\end{figure}

We plot the numerical dispersion relation $\omega$ v.s. $k_1$ (assuming $k_2=0$) in Fig. \ref{fig:1d_num_disp}. We can see from Fig. \ref{fig:1d_num_disp} that, when the order $p$ of $[k_1]$ increases, the dispersion curve is converging to (but never approaches) that of the spectral solver (black solid line). To quantify the locations and growth rates of the NCI modes for high order solvers, in Fig. \ref{fig:nci_16ord_hybrid}, we plot the patterns of the $(\mu,\nu_1)=(0,0)$ and $(0,1)$ NCI modes over the $(k_1,k_2)$ space in the fundamental Brillouin zone. The plot is generated by applying the $p$-th order finite difference operator in $k_1$ and second order finite difference operator in $x_2$ into the theoretical framework developed in Ref. \cite{XuCPC13,YuCPC15}. For completeness we write out the dispersion relation in \ref{sect:app:dispersion}. From Fig. \ref{fig:nci_16ord_hybrid} (a) we can see that the main NCI modes of a high order solver split into two parts: a highly localized  part, i.e., a ``dot'',  near $k_1/k_{g1}=0.2$ (that has a lower growth rate), and another ``strip'' component that is very close to the edge of the fundamental Brillouin zone (that has a higher growth rate). To make both visible on the same scale we multiply the growth rate of the ``dot'' modes by ten. It is interesting to note that the highly localized ``dot'' part of the main NCI modes is located at almost the same place as for the hybrid Yee-FFT solver [shown in Fig. \ref{fig:nci_16ord_hybrid} (c)]. Meanwhile, the ``strip'' component has a growth rate on the same order of magnitude as the main NCI modes of the Yee solver, which are comparable to the $(\mu,\nu_1)=(0,1)$ modes for either FFT or finite difference solvers [see Fig. \ref{fig:nci_16ord_hybrid} (b) and (d)]. This can be explained by the fact that in the low $k_1$ range the dispersion curve of the higher order solver almost overlaps that of the hybrid solver, while for the high $k_1$ range the curve bends down, resulting in a similar NCI pattern to that of the Yee solver (which rolled over for lower $k_1$ values). In the meantime, the $(\mu,\nu_1)=(0,\pm 1)$ modes of the higher order solver reside very close to the edge of the fundamental Brillouin zone [shown in Fig. \ref{fig:nci_16ord_hybrid} (b)], which is similar to the case of the hybrid Yee-FFT solver [Fig. \ref{fig:nci_16ord_hybrid} (d)]. This enables us to readily eliminate these modes by applying a low-pass filter to the current in $k_1$-space.

\begin{figure}[hbtp]
\centering
\includegraphics[width=0.7\textwidth]{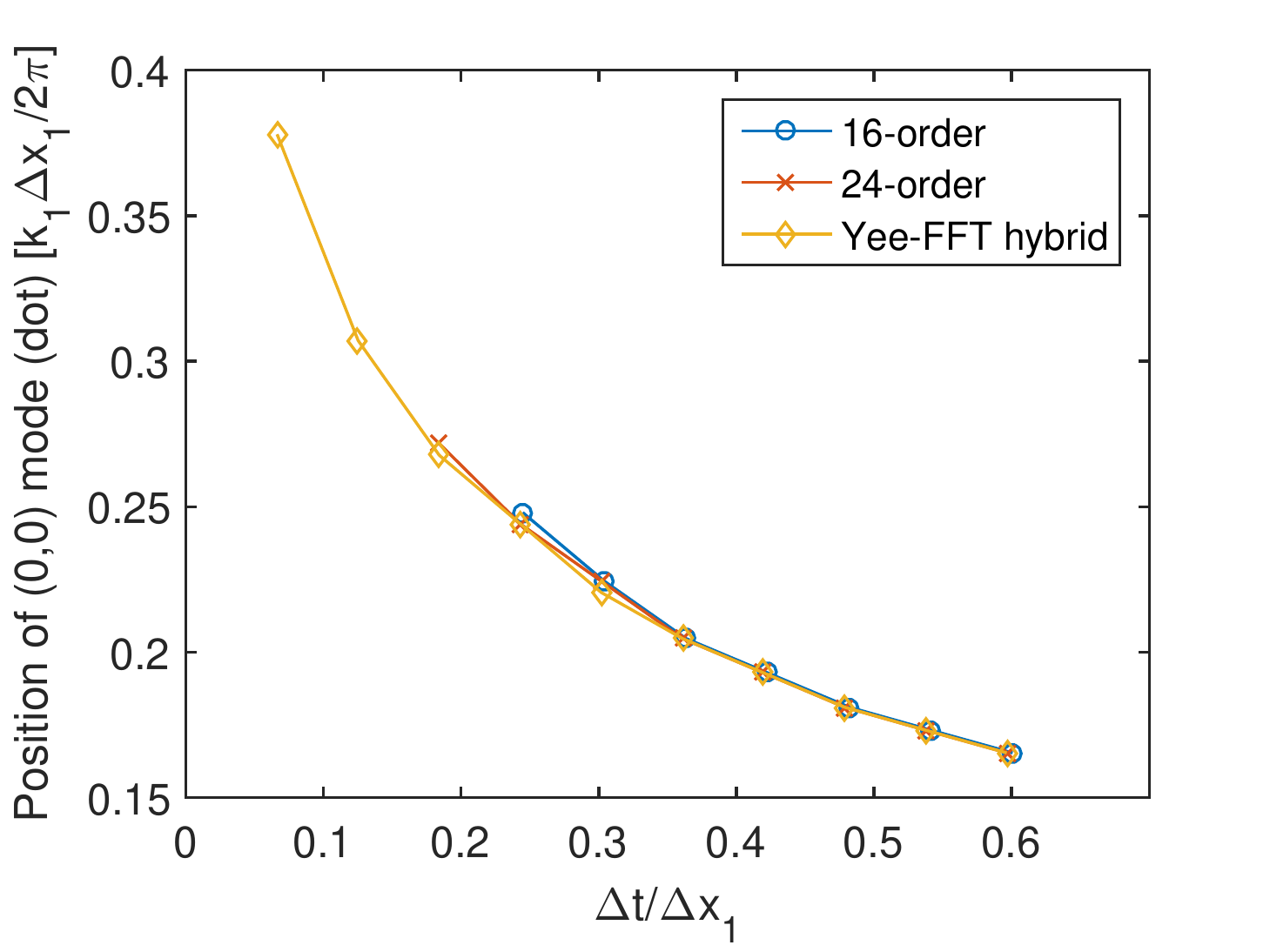}
\caption{The position of the splitted ``dot'' in $(\mu,\nu_1)=(0,0)$ NCI mode at different time step for 16th order, 24th order solver and Yee-FFT hybrid solver. We scan the position using $\Delta x_{1,2}=k_0^{-1}$ and $n_p=n_0$, from $\Delta t=0.1\Delta t_\text{CFL}$ to $\Delta t=0.9\Delta t_\text{CFL}$.}
\label{fig:dot_pos_tstep}
\end{figure}

Just as was the case for the hybrid Yee-FFT solver, the location of the ``dot'' part of the main NCI modes also changes for the higher order solver as one reduces the time step. In Fig. \ref{fig:dot_pos_tstep}, we scan the location of the ``dot'' part of the main NCI modes with different time steps for various solvers.  We can see  that the main NCI modes at moderate $\vert \vec k \vert$ move towards higher  $\vert \vec k \vert$ for both the hybrid Yee-FFT solver, 16th order solver, and 24th order solver. Therefore, it is possible  to apply  the strategies used for the hybrid Yee-FFT solver to effectively eliminate the NCI for the hybrid  higher order-Yee solver. In addition, as $\Delta t/\Delta x_1$ decreases, so do the growth rates for the ``dot'' part of the main NCI modes.

For given simulation parameters, we first calculate the locations of the main NCI modes for the 16th order solver (16th order in $\hat 1$ direction, and 2nd order in other directions). If they are too close to the physical modes, we reduce the time step to move them away from the center towards the edge of the fundamental Brillouin zone. In the next section, we describe how to modify the higher order stencil such that its $k_1$ v.s. $[k_1]_p$ curve has a bump to eliminate the ``dot'' part of the main NCI modes. 

\subsection{Customization of $[k]_1$}
\label{sect:customizeparameter}
In this subsection, we explain how we customize a higher order finite difference first derivative that also has a slight modification near the location of the NCI modes in wave number space. For the FFT-based solvers described in \cite{YuCPC15,YuCPC152} this modification to the EM dispersion relation in the $k_1$ range where the main NCI modes are located can be easily implemented. Specifically, this is accomplished by changing the definition of $k_1$ inside the field solver to $[k_1](k_1)$  in the range $k_1\in [k_{1l},k_{1u}]$, where the $(\mu,\nu_1)=(0,0)$ NCI modes reside to $[k_1]=k_1+\Delta k_\text{mod}(k_1)$ where $\Delta k_\text{mod}(k_1)$ is a small localized perturbation(see Fig. 3(a) in \cite{YuCPC152}). It usually takes the form of 
\begin{equation}\label{eq:dk1}
	\Delta k_\text{mod}(k_1)=
	\begin{cases}
	\Delta k_\text{mod,max}\sin\left(\pi\frac{k_1-k_{1l}}{k_{1u}-k_{1l}}\right)^2, &k_{1l} \le k_1 \le k_{1u} \\
	0, &\text{otherwise}
	\end{cases}
\end{equation}
where $k_{1l},k_{1u}$ are the lower and upper bounds of the region that is modified, and $\Delta k_\text{mod,max}$ is the maximum of $\Delta k_\text{mod}$. In an FFT-based solver this modification is straightforward to implement, while in a high order FDTD solver, one has to find a stencil that has both higher order accuracy for the derivative over a wide range of wave number space as well as a modification in a local region of wave number space. For a regular $p$th order solver (where $p$ is an even number), there are $p/2$ coefficients, $C_l^p$, for the stencil and the numerical dispersion relation is uniquely determined. It naturally follows that if we want to customize the dispersion relation based on the high order solver, we will need to add more degrees of freedom, i.e., more coefficients, into the operator. This means the stencil will be broader. The coefficients will still need to be constrained so that the operator has higher-order accuracy, while at the same time it has the desired modification in a local region of $k_1$ space.

We denote the high order solver as $\nabla^\pm_{p*}=\left( \partial_{p*,x_1}^\pm, \partial_{2,x_2}^\pm, \right)$. The first component has the form
\begin{eqnarray}
	\partial_{p*,x_1}^+f_{i_1,i_2} &=& \frac{1}{\Delta x_1}\sum_{l=1}^M \tilde{C}_l^p \left(f_{i_1+l,i_2}-f_{i_1-l+1,i_2}\right) \\
	\partial_{p*,x_1}^-f_{i_1,i_2} &=& \frac{1}{\Delta x_1}\sum_{l=1}^M \tilde{C}_l^p \left(f_{i_1+l-1,i_2}-f_{i_1-l,i_2}\right)
\end{eqnarray}
while the second component  is still the standard second order accurate operator. The modified  solver has $M$ coefficients, $\tilde{C}^p_l$, where $M>p/2$. The corresponding finite difference operator in $k$-space becomes
\begin{equation}\label{eq:k1customize}
	[k_1]_{p*}=\sum_{l=1}^M \tilde{C}_l^p \frac{\sin[(2l-1)k_1\Delta x_1/2]}{\Delta x_1/2}
\end{equation}
In order to construct the ``bump'' in the dispersion curve for the proposed solver, we need to find an ``optimized'' set of $\tilde{C}^p_l$ such that $[k_1]_{p*}$ will best approximate the modified $[k_1] = [k_1]_p+\Delta k$ described in Eq. (\ref{eq:dk1}). For the purpose of simplifying the notation, in the following we normalize $[k_1]_{p*}$, $[k_1]_p$, $\Delta k_\text{mod}$, $k_{1l}$, $k_{1u}$ and $k_1$, with $k_{g1}=2\pi/\Delta x_1$. In the spirit of the least square approximation, we construct a function $F_1$ 
\begin{equation}
	F_1=\int_0^{1/2}\left( [k_1]_{p*}-[k_1]_{p}-\Delta k_\text{mod} \right)^2 dk_1
\end{equation}
which we will minimize to find an optimum  set of $\tilde{C}_j^p$. We note that a weight function $w(k_1)$ can be employed and this is an area for future work. In addition, the high order solver should also meet the requirement of $p$th order accuracy and satisfy the condition $\partial^\pm_{p*,x_1}\rightarrow \partial_{x_1}$ as $\Delta x_1 \rightarrow 0$. Therefore, the coefficients should be subject to the linear equations below
\begin{equation}
	\mathcal{M} \tilde{\vec{C}}^p=\hat{\vec{e}}_1
\end{equation}
where $\tilde{\vec{C}}^p=(\tilde{C}_1^p,...,\tilde{C}_M^p)^T$, $\hat{\vec{e}}_1=(1,0,...,0)^T$ and the elements of the matrix $\mathcal M$ are $\mathcal M_{ij}=(2j-1)^{2i-1}/(2i-1)!\ (i=1,...,p/2)$ and $(j=1,...,M)$. This is a constrained least-square minimization problem so we can use the Lagrange multipliers to solve it. 

The Lagrangian is defined by $\mathcal{L}=F_1+F_2$ where $F_2=\vec{\lambda}^T(\mathcal{M}\tilde{\vec{C}}^p-\hat{\vec{e}}_1)$ and $\vec{\lambda}=(\lambda_1,...,\lambda_{p/2})^T$. Solving the constrained least-square minimization problem is equivalent to solving,
\begin{equation}
\label{eq:lagrange}
\partial\mathcal{L}/\partial \tilde{C}_j^p=0\ (j=1,...,M)\ \text{and}\ \partial\mathcal{L}/\partial \lambda_i=0\ (i=1,...,p/2)
\end{equation}
It can be straightforward to show that this results in the following set of equations, 
\begin{eqnarray}
\label{eq:lag1}
\frac{\partial F_1}{\partial \tilde{C}_j^p} &=&
\begin{cases}
	\frac{1}{2\pi^2}(\tilde{C}_j^p-C_j^p-A_j), & 1 \le j \le \frac{p}{2} \\
	\frac{1}{2\pi^2}(\tilde{C}_j^p-A_j), & \frac{p}{2}+1 \le j \le M
\end{cases} \\
\label{eq:lag2}
\frac{\partial F_2}{\partial \tilde{C}_j^p} &=& \sum_i \lambda_i \mathcal M_{ij} \\
\label{eq:lag3}
\frac{\partial F_2}{\partial \lambda_j} &=& \sum_i \mathcal M_{ji}A_i-e_j 
\end{eqnarray}
where
\begin{equation}
	A_j=\frac{8\Delta k_\text{mod,max}(\cos[(2j-1)\pi k_{1u}]-\cos[(2j-1)\pi k_{1l}])}{(2j-1)[(2j-1)^2(k_{1u}-k_{1l})^2-4]}
\end{equation}
Combing Eqs. (\ref{eq:lag1})-(\ref{eq:lag3}), we can reformat Eq. (\ref{eq:lagrange}) into a matrix equation
\begin{equation}
\label{eq:lagrange_matrix}
	\begin{pmatrix} 
	\frac{1}{2\pi^2}\vec{I} & \mathcal{M}^T \\ 
	\mathcal{M} & \vec{0} 
	\end{pmatrix}
	\begin{pmatrix} \tilde{\vec{C}}^p \\ \vec{\lambda} \end{pmatrix} =
	\begin{pmatrix} \frac{1}{2\pi^2}(\vec{A}+\vec{C}^p) \\ \hat{\vec{e}}_1 \end{pmatrix}
\end{equation}
where $\vec{I}$ is the $M\times M$ unit matrix and $\vec{C}^p=(C_1^p,...,C_{p/2}^p,0,...,0)^T$. For given ``bump'' parameters $\Delta k_\text{mod,max}$, $k_{1l}$ and $k_{1u}$, Eq. (\ref{eq:lagrange_matrix}) can be solved numerically. Henceforth, in this paper we use $M=p$.

\begin{figure}[hbtp]
\centering
\includegraphics[width=0.95\textwidth]{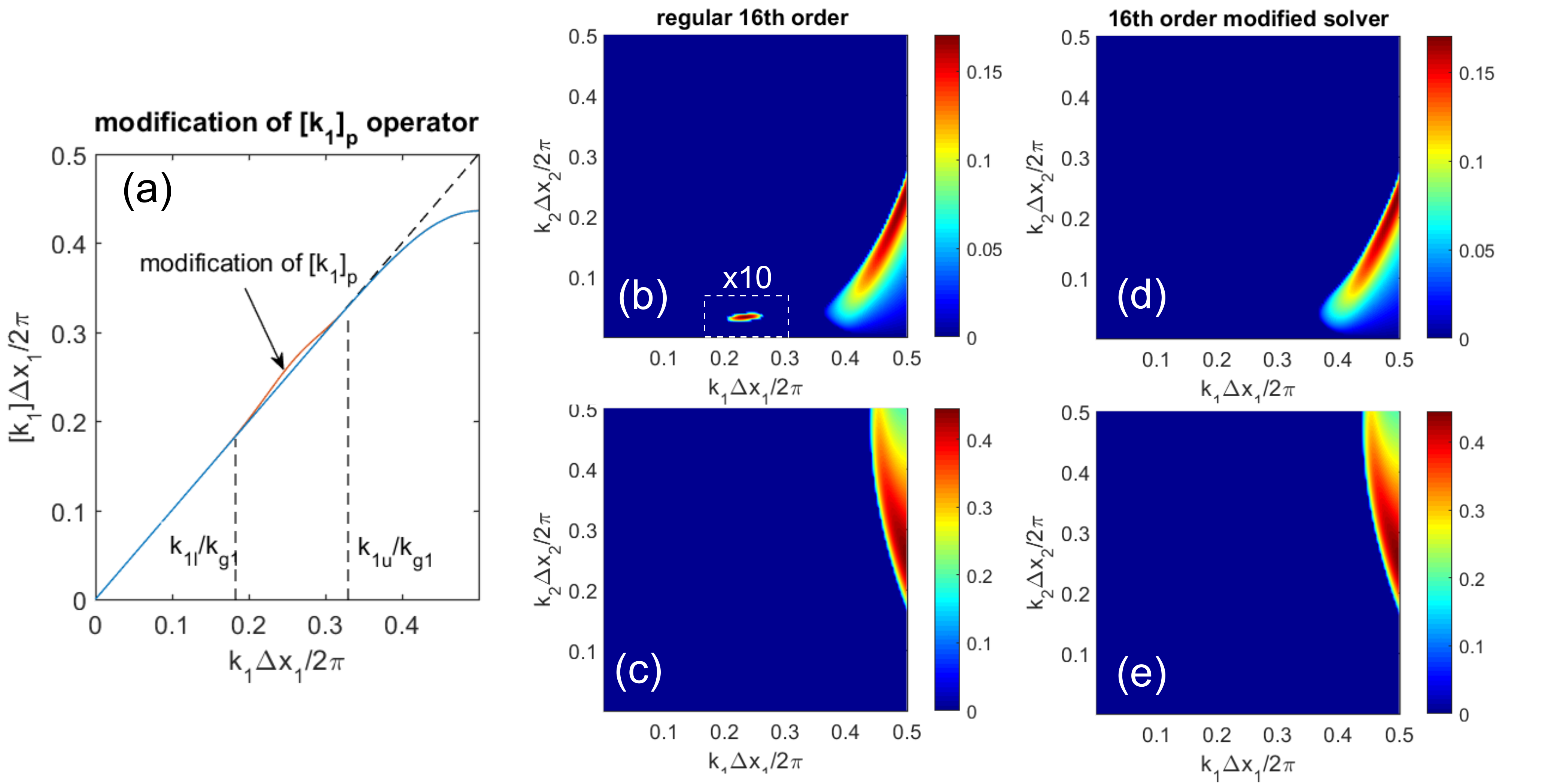}
\caption{In (a) the perturbation (red line) introduced by the modified 16th order solver is shown. The blue line denotes the $[k_1]\text{-}k_1$ relation of the regular 16th order solver. The lower and upper limits of perturbation are $k_{1l}/k_{g1}=0.18$ and $k_{1u}/k_{g1}=0.33$. The perturbation magnitude $\Delta k_\text{mod,max}=0.01$. (b) and (c) are $(\mu,\nu_1)=(0,0)$ and $(\mu,\nu_1)=(0,1)$ NCI mode patterns of the regular 16th order solver respectively. The values in the dashed line box in (b) is multiplied by 10 for better visualization. (d) and (e) are the patterns of the modified 16th order solver. When generating these plots we use $\Delta x_{1,2}=0.2k_0^{-1}$, $\Delta t=0.05\omega_0^{-1}$ and $n_p=50n_0$}
\label{fig:bump}
\end{figure}

In Fig. \ref{fig:bump} we show the comparison of the$[k_1]$ operator and the NCI mode patterns between the regular and customized high order solver. We use a 16th order solver as an example. In Fig. \ref{fig:bump}(a), we show that a perturbation (red line) can be introduced to the $[k_1]_{16}$ operator within the bump region, while the operators $[k_1]_{16}$ and $[k_1]_{16*}$ are almost overlapped outside the region. Fig. \ref{fig:bump}(b)(c) show the NCI mode patterns for the regular 16th order solver. We can see the dot of $(\mu,\nu_1)=(0,0)$ modes presented in the middle of the fundamental Brillouin zone, for which we aim to eliminate through the use of the modified solver. In Fig. \ref{fig:bump}(d), the dot has been eliminated and $(\mu,\nu_1)=(0,1)$ modes and the remaining of $(\mu,\nu_1)=(0,0)$ modes are almost identical to what was seen for the regular high order solver.

\subsection{Courant condition}
The derivation of the Courant condition for the proposed high order solver is straightforward. From the numerical dispersion relation
\begin{equation}
	[\omega]^2 = c^2\left([k_1]^2_{p*}+[k_2]^2_2\right),
\end{equation}
it can be shown that to keep $\omega$ a real number, the corresponding constraint on the time step must be satisfied, \emph{i.e.}
\begin{equation}
	\frac{\Delta t}{2}\sqrt{\left(\sum_{l=1}^M \tilde{C}_l^p \frac{\sin[(2l-1)k_1\Delta x_1/2]}{\Delta x_1/2}\right)^2 + \left(\frac{\sin(k_2\Delta x_2/2)}{\Delta x_2/2}\right)^2} \le 1.
\end{equation}
Note that $|k_1|\le\pi/\Delta x_1$ and $|k_2|\le\pi/\Delta x_2$, we obtain the Courant condition of the proposed high order solver
\begin{equation}
	\Delta t \le 1/\sqrt{\frac{\left(\sum_{l=1}^M\tilde{C}^p_l\right)^2}{\Delta x_1^2} + \frac{1}{\Delta x_2^2}}
\end{equation}

For the standard high order solver, given the cell sizes, the Courant limit only depends on cell sizes and the order solver's accuracy. For instance, the Courant limit of a 16th order solver (16th order in $x_1$, while second order in $x_2$) is $\Delta t_{CL}= 0.6575\Delta x_1$ with $\Delta x_1=\Delta x_2$. As for the customized solver, although the specified solver coefficients $\tilde{C}_l^p$ depend on the modification of numerical dispersion we introduce, the Courant limit on the time step varies little as we alter the ``bumps'' in the numerical dispersion curve. Taking the 16th order customized solver with 16 coefficients and $\Delta x_1=\Delta x_2$ for example, the Courant condition reduces to $\Delta t_{CL}= 0.6550\Delta x_1$ for $k_l=0.1$, $k_u=0.3$, $\Delta k_\text{mod}=0.01$, and $\Delta t_{CL}= 0.6562\Delta x_1$ for $k_l=0.15,k_u=0.3,\Delta k_\text{mod}=0.005$. As we can see, the Courant condition changes little when switching from the standard high order solver to the customized solver.

\subsection{Cartesian 3D and quasi-3D scenarios}
As can be seen from previous sections, this FDTD solver only modifies the finite difference operator in the plasma drifting direction. As a result, although not presented in this paper, the method described above can be extended to Cartesian 3D and quasi-3D geometry \cite{quasi3d,davidson} 
in a straightforward way. 

\section{Charge conservation and parallelization of the solver}
\label{sect:current}

Similar to the case in hybrid Yee-FFT solver, when the $[k_1]$ of the solver is different from the second order accurate $[k_1]_2$, one needs to apply a correction to the current in order to satisfy Gauss' Law. This is due to the fact that in a typical FDTD EM-PIC code, the EM fields are advanced by the Faraday's law and Ampere's law, while the Gauss' Law is satisfied by applying a charge conserving current deposit \cite{chargeconservation}. The charge conserving current deposition is second order accurate in all directions, which matches exactly to the standard Yee solver. As a result, when the finite difference operator for the derivative along a particular direction changes in a solver, Gauss' Law is no longer satisfied if the current is not corrected correspondingly.

More specifically, the charge conserving current deposition ensures the second-order-accurate finite difference representation of the continuity equation,
\begin{equation}
	\overline{\frac{\partial}{\partial t}} \rho^n+\nabla^-_2\cdot\vec{J}^{n+\frac{1}{2}}=0
\end{equation}
where $\overline{\frac{\partial}{\partial t}}G^n=\frac{G^{n+1}-G^n}{\Delta t}$ for an arbitrary scalar quantity $G^n$. For the Yee solver, Gauss's law is rigorously satisfied every time step if it is satisfied at the beginning. However, when combining the high order solver and the second-order-accurate current deposition scheme, we need to apply a correction to the current in the drifting direction in order that Gauss's law remains satisfied throughout the whole simulation. After the current has been calculated locally, we then ``correct'' them by performing an FFT along the $x_1$ direction,
\begin{equation}
	\label{eq:current_corr}
	\tilde{J}_1^{n+\frac{1}{2}}=\frac{[k_1]_2}{[k_1]_{p*}}J_1^{n+\frac{1}{2}}
\end{equation}
where $\tilde{J}_1$ is the corrected current, in a similar manner to what was employed for the hybrid Yee-FFT solver. Performing Fourier transform in the $x_1$ direction and applying the correction scheme in Eq. (\ref{eq:current_corr}), we guarantee
\begin{equation}
	\label{eq:con_k}
	\overline{\frac{\partial}{\partial t}} \mathcal{\rho}^n(k_1,x_2)+i[k_1]_{p*}\tilde J_1^{n+\frac{1}{2}}(k_1,x_2)+\partial^-_{2,x_2}J_2^{n+\frac{1}{2}}(k_1,x_2)=0
\end{equation}
Combining Eq. (\ref{eq:con_k}) with Ampere's law, Eq. (\ref{eq:maxwell_eq2}), (replacing $\nabla^-_p$ with $\nabla^-_{p*}$ to be consistent with the modified high order solver), we obtain
\begin{equation}
	\overline{\frac{\partial}{\partial t}}\left(-4\pi\rho^n+i[k_1]_{p*}E_1^n+\partial^-_{2,x_2}E_2^n\right)=0
\end{equation}
We carry out inverse Fourier transform to retrieve the equation in real space,
\begin{equation}
	\overline{\frac{\partial}{\partial t}}\left(-4\pi\rho^n+\nabla^-_{p*}\vec{E}^n\right)=0
\end{equation}
which indicates that Gauss's law is satisfied if it is satisfied initially.

It is important to note that one effect that arises from the current correction is that the current from one particle extends to more cells. Therefore, an originally localized current distribution would be spread out over more cells after we correct the current in the $k$-space and transform back to real space. This results from the use of a less local operator for the derivative. Nevertheless, the current still rigorously satisfies the continuity equation for the desired particle shape. We point out that this effect can be neglected and the current correction is still a nearly error-free scheme in some sense, as will be discussed below.

Assuming we have a point current initially in the real space, located at the grid index $i_1=0$, \emph{i.e.} $J_{1,i_1}=\delta(i_1)$. After performing the discrete Fourier transform, the current in the $k$-space becomes unity for all $k_1$, \emph{i.e.} $J_{1,\kappa_1}=1$, where $\kappa_1=-N/2,...,N/2-1$ is the mode number, and $k_1=2\pi\kappa_1/N\Delta x_1$ with $N$ the number of cells in $x_1$ direction. The corrected current $\tilde{J}_{1,\kappa_1}$ is therefore  the correction factor
\begin{equation}
\label{eq:corr_fac}
	\left(\frac{[k_1]_2}{[k_1]_{p*}}\right)_{\kappa_1}=\frac{\sin\left(\frac{\pi}{N}\kappa_1\right)}{\sum_l \tilde{C}^p_l \sin\left[\frac{(2l-1)\pi}{N}\kappa_1\right]}
\end{equation}
It can be asserted that the inverse discrete Fourier transform (IDFT) of the correction factor extends over all space. This can be easily proved by contradiction. Suppose the IDFT of the corrected point current was confined to a region of space, e.g., it is non zero only in the interval $[-W,W]$ and the expanded current distribution in the real space is symmetric, \emph{i.e.} $J_{1,i_1}=J_{1,-i_1}$. The corresponding Fourier coefficients after a DFT would therefore be a finite summation of cosine functions, \emph{i.e.} $J_{1,\kappa_1}=J_{1,i_1=0}+2\sum_{i_1=1}^W J_{1,i_1}\cos(2\pi i_1\kappa_1/N)$. By inspection, it is clear that, Eq. (\ref{eq:corr_fac}) cannot be rewritten in such way. Therefore, we have proven that the IDFT of the current correction for a delta function cannot be localized in space and therefore, must exist within the entire space over which the FFT is performed.

\begin{figure}[hbtp]
\centering
\includegraphics[width=0.8\textwidth]{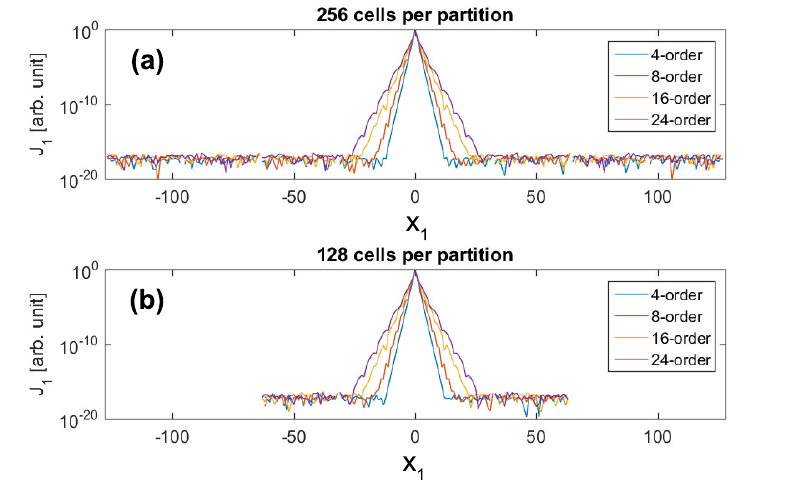}
\caption{Effect of the current expansion tested by point current. Numerical calculations are carried out on 256 cells (a) and 128 cells (b), to model the cases using different partition sizes. Current with Dirac-delta distribution is initialized and the current corrections of different orders of solver are applied in the $k$-space. (a) and (b) show the current distributions in real space with different correction schemes. We set $\Delta x_1=1$ for the calculations.}
\label{fig:current_expansion}
\end{figure}

However, as we show next the values of the corrected current for an initial delta function (at the origin) fall off rapidly for cells away from the origin. In fact, the values fall below double precision accuracy effectively making the corrected current effectively only non-zero in  finite region. We illustrate this through 1D numerical calculations. We initialize the current as a Dirac-delta distribution in the center of the grid, \emph{i.e.} $J_1=\delta(x_0)$. Then we perform a 1D FFT to the current, then use the correction according to Eq. (\ref{eq:current_corr}) for solvers with different orders in the $k$-space,  and finally perform an IFFT to transform the current back into real space. Fig. \ref{fig:current_expansion} shows the spatial distribution of the corrected current. It can be seen that the extent of the current is widened by the correction process and the width increase as the order of the solver is increased. Nonetheless, the expansion remains in a localized region in real space and beyond this region the amplitude is on the order of  $10^{-16}$, which corresponds to double precision roundoff. The fact that the current is localized indicates that the current correction can be done on a local domain so long as a sufficient number of guard cells are used. The permits of using domain decomposition along the $\hat 1$ direction and the use of a local FFT on each domain. For example, we have decreased the size of the domain from 256 grids in Fig. \ref{fig:current_expansion} (a) to 128 grids in Fig. \ref{fig:current_expansion} (b). The current distribution, the width expansion, and noise level are almost the same. Therefore, we can see that the size of a parallel partition makes little impact on the simulation results. Although the current expansion is mathematically infinite as mentioned before, in the sense of considering the precision of the numerical algorithm, the current expansion can be viewed as localized. When a sufficient number of guard cells for each simulation partition is used, this current correction scheme will be nearly free of any error brought by the current expansion.

\section{Sample simulations}
\label{sect:simulation}

In this section we present sample simulations using the customized solver and its corresponding NCI elimination schemes. In these simulations, we used the low-pass filtering for current in the form of 
\begin{equation}
	F(k_1)=
	\begin{cases}
		1, & |k_1|<f_l k_{g1} \\
		\sin^2\left(\frac{k_1-f_u k_{g1}}{f_l k_{g1}-f_u k_{g1}}\frac{\pi}{2}\right), & f_l k_{g1}\le k_1 \le f_u k_{g1}\\
		0, & |k_1|>f_u k_{g1}
	\end{cases}
\end{equation}
The filter retains the $k_1$ modes smaller than $f_l k_{g1}$ and cuts off the modes larger than $f_u k_{g1}$. A $\sin^2$ function is used between $f_l k_{g1}$ and $f_u k_{g1}$ for smooth connection between unity and zero. 

\subsection{Drifting plasma}
\label{sect:simulation:drift}
In this subsection we demonstrate the NCI elimination capability of the customized FDTD Maxwell solver with a 2D Cartesian OSIRIS simulation of drifting plasma. We fill the simulation box with a plasma drifting relativistically at $\gamma = 20$ along $\hat 1$ direction. The plasma has a very small but finite temperature to seed the NCI. Periodic boundary conditions are used for both the $\hat{1}$ and $\hat{2}$ directions. We performed simulations using both the 16th order solver, the customized solver, and the hybrid Yee-FFT solver, with and without the low-pass filters. Other simulation parameters are presented in Table \ref{tab:driftingplasmapara}, and the corresponding coefficients for the customized solver are listed in Table \ref{tab:customizepara1}. 

\begin{table}[hbtp]
\centering
\begin{tabular}{p{6cm}c}
\hline\hline
\textbf{Parameters} & \textbf{Values}\\
\hline
grid size $(\Delta x_{1},\Delta x_{2})$ & $(0.5k^{-1}_0, 0.5k^{-1}_0)$\\
time step $\Delta t$ & $0.25\Delta x_1$\\
number of grid & $512\times 512$ \\
particle shape & quadratic, cubic\\
electron drifting momentum $p_{10}$ &19.975 $m_ec$\\
plasma density & 2.0$~n_0$\\
\hline\hline
\end{tabular}
\caption{Simulation parameters for the 2D drifting plasma simulation. $n_p$ is the plasma density, and $\omega^2_0=4\pi q^2n_0/m_e$, $k_0=\omega_0$ ($c$ is normalized to 1).}
\label{tab:driftingplasmapara}
\end{table}
\begin{table}[hbtp]
\centering
\begin{tabular}{cc|cc}
\hline\hline
\textbf{Coefficients} & \textbf{Values} & \textbf{Coefficients} & \textbf{Values}\\
\hline
$\tilde C^{16}_{1}$ &1.237042976225048 & $\tilde C^{16}_{2}$ & -0.102548201854464\\
$\tilde C^{16}_{3}$ &0.022015354460742 & $\tilde C^{16}_{4}$ & -0.009258452621442\\
$\tilde C^{16}_{5}$ &0.000410036656959 & $\tilde C^{16}_{6}$ & 0.002572239519500\\
$\tilde C^{16}_{7}$ &0.001482836071727 & $\tilde C^{16}_{8}$ &-0.001392055950412 \\
$\tilde C^{16}_{9}$ &-0.001472515326959 & $\tilde C^{16}_{10}$ &0.000478783514362 \\
$\tilde C^{16}_{11}$ &0.001200462462019 & $\tilde C^{16}_{12}$ & -0.000187062256742\\
$\tilde C^{16}_{13}$ &-0.001059471474041 & $\tilde C^{16}_{14}$ & 0.000873314953435\\
$\tilde C^{16}_{15}$ &-0.000281855449164 & $\tilde C^{16}_{16}$ & 0.000034281167855\\
\hline\hline
\end{tabular}
\caption{Coefficients $\tilde C^{16}_{i}$ in Eq. (\ref{eq:k1customize}) for the customized solver based on the 16th order solver, for the single plasma drift simulation discussed in section \ref{sect:simulation:drift}.}
\label{tab:customizepara1}
\end{table}

\begin{figure}[hbtp]
\centering
\includegraphics[width=0.8\textwidth]{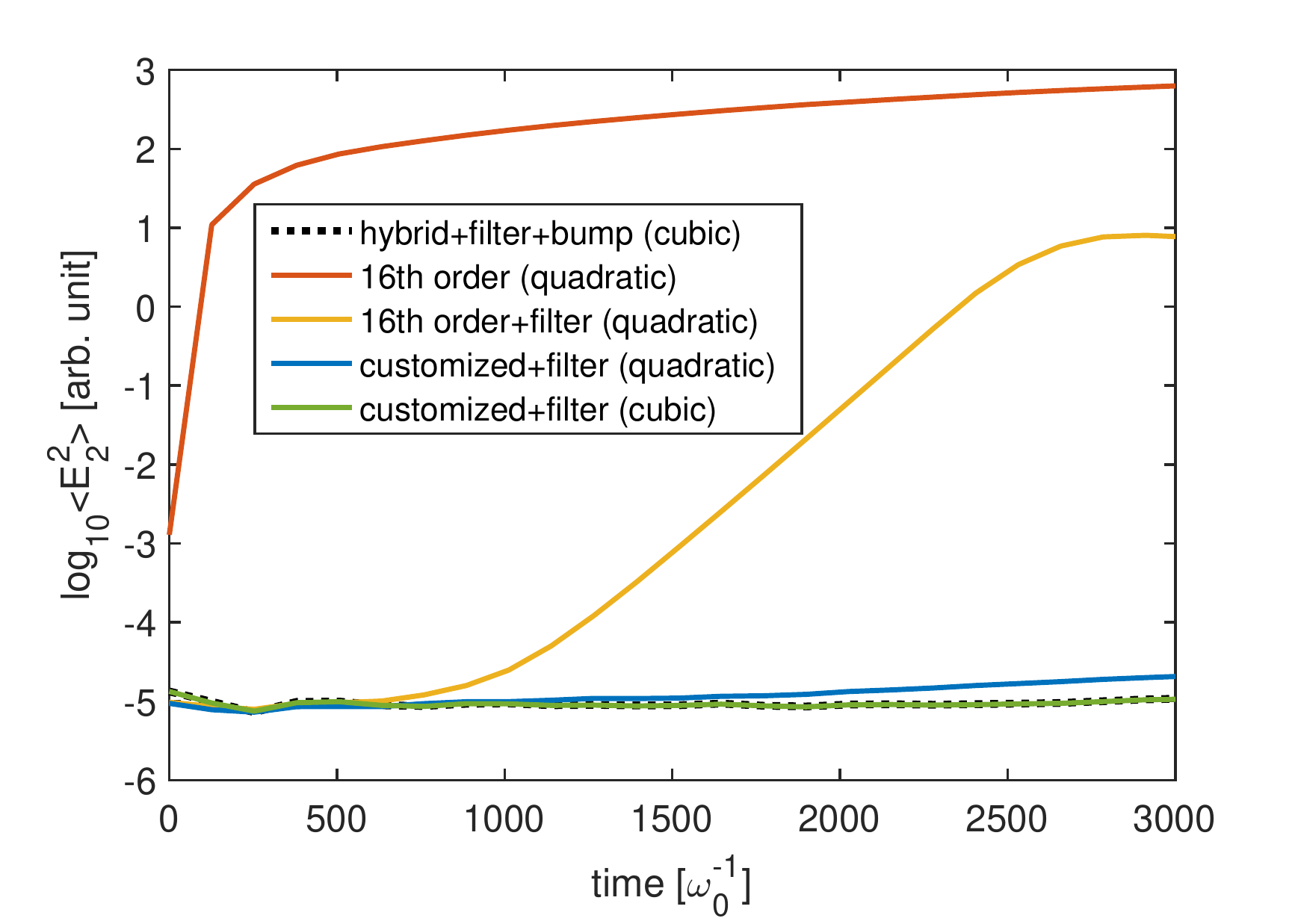}
\caption{Evolutions of the $E_2$ energies in the numerical systems for various setups in drifting plasma 2D Cartesian PIC simulations, as discussed in section \ref{sect:simulation:drift}.}
\label{fig:sim_drift}
\end{figure}

We can see from Fig. \ref{fig:sim_drift} that, by comparing the red line (16th order solver without any filters) and the orange line (16th order solver plus low-pass filter), that applying the low-pass filter to a 16th order solver significantly reduces the growth of the $E_2$ energy. This is because the fastest growing $(\mu,\nu_1)=(0,\pm 1)$ NCI modes are eliminated by the low-pass filter. Besides using the low-pass filter, when we add the bump to the 16th order solver (thus making it a customized solver), the growth rate is further reduced since the main NCI modes are completely eliminated (see the blue line). Even higher order NCI modes \cite{YuCPC15} are attributed to the slight growth in energy for this case (blue line in Fig. \ref{fig:sim_drift}), and when the cubic particle shape is applied, the corresponding energy growth is effectively suppressed (green line in Fig. \ref{fig:sim_drift}). We can see that when the low-pass filter, bump, and higher order particles are applied to the cases of both customized solver (green line) and hybrid Yee-FFT solver (black dotted line), both the hybrid Yee-FFT solver and customized solver schemes effectively eliminate the NCI. 

\subsection{Relativistic shock}
\label{sect:simulation:shock}
We next present an example where two plasmas are collided against each other which is relevant for relativistic shock simulations. The two plasmas drift towards each other with a Lorentz factor of $\gamma=20.0$. The simulation has a box size of $131072\times 2048$, for which the number of cells in $\hat 1$ direction is much larger than that of the $\hat 2$ direction. Since the plasmas are drifting in the $\hat 1$ direction, the total number of cores that can be used in such a simulation would be significantly limited if we use FFT-based solvers, which requires one partition along the $\hat 1$ direction. With the customized solver and corresponding elimination scheme, we are able to partition in the $\hat 1$ direction. We used a 2D domain decomposition with  $256\times 16$ partitions along the $\hat 1$ and $\hat 2$ directions respectively.  Other simulation parameters are listed in Table \ref{tab:shock}, and the corresponding coefficients for the customized solver are listed in Table \ref{tab:customizepara2}. 

\begin{figure}[hbtp]
\centering
\includegraphics[width=1\textwidth]{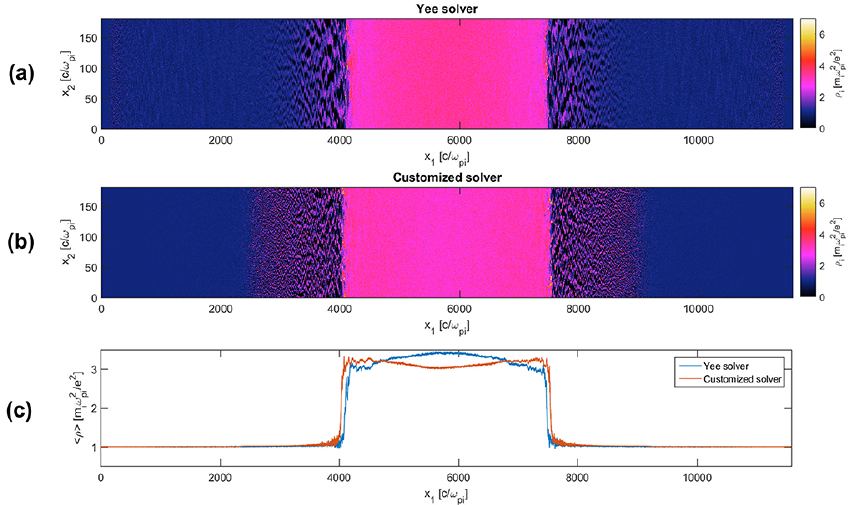}
\caption{The ion densities and their line outs for a relativistic shock simulation, as discussed in section \ref{sect:simulation:shock}. The corresponding simulation parameters are listed in Table \ref{tab:shock}.}
\label{fig:sim_shock}
\end{figure}

In Fig. \ref{fig:sim_shock} we plot the 2D color isosurface plots of the ion density, and line outs of the $x_2$ averaged ion density for the Yee solver, and customized solver. For the Yee solver case, we used the optimal time step of $\Delta t=0.5 \Delta x_1$ at which the NCI is minimized, plus a 5-pass current smoothing and compensation for the current, and the EM fields are also filtered every two time steps. For the case with the customized solver, we used the same NCI elimination scheme as is used for the single drifting plasma case, but with slightly different coefficients (different density and time step) for the solver. We can see from Fig. \ref{fig:sim_shock} that there are noticeable differences. For example, the transverse size of the filaments in the density are larger for the Yee case than for the customized case. From our NCI theory we know that the Yee solver with the optimized time step does not eliminate the main $(\mu,\nu_1)=(0,0)$ modes. The growth rate for these modes is reduced, but they are not localized in space; instead they reside within the range of physics. It is not obvious when, and how these modes are altering the physics. On the other hand,  the customized solver together with our filters completely removes the $(\mu,\nu_1)=(0,\pm 1)$ and $(\mu,\nu_1)=(0,0)$ modes; and the use of the higher order particle shapes reduces the growth rate for the next highest growing modes. 

\begin{table}[hbtp]
\centering
\begin{tabular}{p{10cm}c}
\hline\hline
\textbf{Parameters} & \textbf{Values}\\
\hline
Plasma  &  \\
\quad density $n_{pe},n_{pi}$ & $n_0$ \\
\quad initial Lorentz factor $\gamma_0$ & 20.0  \\
\quad initial thermal velocity $v_{\text{th},e,i}$ & $8.7\times10^{-5}c$ \\
\quad mass ratio $m_i/m_e$ &  32 \\
\multicolumn{2}{l}{Simulation using customized high order solver}  \\
\quad cell size $\Delta x_{1,2}$ & 0.5$k_0^{-1}$ \\
\quad time step $\Delta t/\Delta x_1$ & 0.2 \\
\quad number of cells & $2^{17}\times2^{11}$ \\
\quad particle shape & cubic \\
\quad particle per cell & $(1,2)$ \\
\quad $[k_1]$ modification $(k_{1l}, k_{1u}, \Delta k_\text{mod,max})$ & $(0.1,0.35,0.01)$ \\
\quad lowpass filter $(f_l,f_u)$ & (0.275,0.3) \\
\multicolumn{2}{l}{Simulation using standard Yee solver}  \\
\quad cell size $\Delta x_{1,2}$ & 0.5$k_0^{-1}$ \\
\quad time step $\Delta t/\Delta x_1$ & 0.5 \\
\quad number of cells & $2^{17}\times2^{11}$ \\
\quad particle shape & cubic \\
\quad particle per cell & $(1,2)$ \\
\hline\hline
\end{tabular}
\caption{Parameters for 2D relativistic collisionless plasma simulations in lab frame using the modified high order solver and Yee solver. The plasma density $n_0$ and corresponding wave number $k_0$ are used to normalize the simulation parameters. The parameters of $[k_1]$ modification are normalized to $k_{g1}\equiv 2\pi/\Delta x_1$.}
\label{tab:shock}
\end{table}

\begin{table}[hbtp]
\centering
\begin{tabular}{cc|cc}
\hline\hline
\textbf{Coefficients} & \textbf{Values} & \textbf{Coefficients} & \textbf{Values}\\
\hline
$\tilde C^{16}_{1}$ & 1.243205632406442 & $\tilde C^{16}_{2}$ & -0.096527073844747 \\
$\tilde C^{16}_{3}$ & 0.017018941335700 & $\tilde C^{16}_{4}$ & -0.013839950216042 \\
$\tilde C^{16}_{5}$ &  0.003588768352855 & $\tilde C^{16}_{6}$ & 0.005153133591937\\
$\tilde C^{16}_{7}$ &  0.000007068893273 & $\tilde C^{16}_{8}$ &  -0.002317133408538\\
$\tilde C^{16}_{9}$ & -0.001166192174494 & $\tilde C^{16}_{10}$ & 0.000552266782136\\
$\tilde C^{16}_{11}$ & 0.001508596910066 & $\tilde C^{16}_{12}$ &  -0.000134050410326\\
$\tilde C^{16}_{13}$ & -0.001599956501178 & $\tilde C^{16}_{14}$ & 0.001305552125425\\
$\tilde C^{16}_{15}$ &-0.000423469804615 & $\tilde C^{16}_{16}$ &0.000051829248350\\
\hline\hline
\end{tabular}
\caption{Coefficients $\tilde C^{16}_{i}$ in Eq. (\ref{eq:k1customize}) for the customized solver based on the 16th order solver, for the relativistic shock simulations, and LWFA simulations in the Lorentz boosted frame, as discussed in section \ref{sect:simulation:shock} and section \ref{sect:simulation:lwfa}.}
\label{tab:customizepara2}
\end{table}

\subsection{LWFA boosted frame simulation}
\label{sect:simulation:lwfa}
In this subsection, we present 3D Cartesian LWFA boosted frame PIC simulations using the customized FDTD Maxwell solver in OSIRIS. For comparison, we carried out simulations using the hybrid Yee-FFT solver and customized FDTD Maxwell solver respectively. The corresponding lab frame simulation is the 1.3 GeV case discussed in \cite{LuScaling}, and we have also listed the parameters in Table \ref{tab:lwfa_pars}. Note that although the simulation parameters in this scenario is different from those of the relativistic shock simulations discussed in section \ref{sect:simulation:shock}, the locations of the main NCI modes for a 16th order solver under these two sets of parameters are very close to each other. Therefore we used the same coefficients for the customized solver as in section \ref{sect:simulation:shock}, as listed in Table \ref{tab:customizepara2}. 

\begin{figure}[hbtp]
\centering
\includegraphics[width=0.9\textwidth]{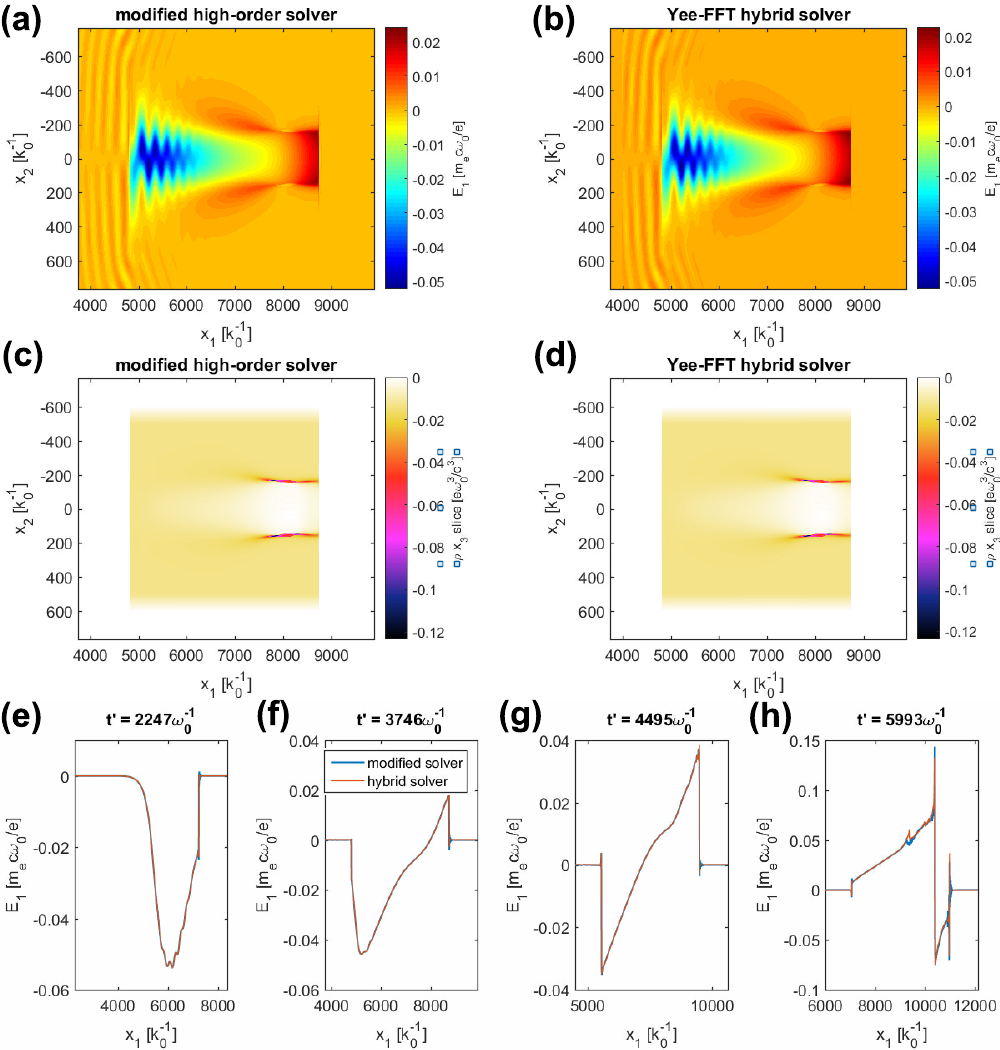}
\caption{Comparison of simulations in the boosted frame between the customized high order solver and Yee-FFT hybrid solver. 2D plots of $E_1$ field at $t'=3746~\omega_0^{-1}$ for both solvers are shown in (a) and (b). The electron density profiles are shown in (c) and (d). (e) to (h) plot the on-axis lineouts of $E_1$ fields at different times.}
\label{fig:sim_lwfa}
\end{figure}

In Fig. \ref{fig:sim_lwfa}(a) and (b) we plot the $E_1$ field at $t'=3746~\omega_0^{-1}$ for simulations with either a modified high order solver or a hybrid Yee-FFT solver. Both solvers give nearly identical results and no evident numerical Cerenkov radiation is observed in either cases. In Fig. \ref{fig:sim_lwfa} (c) and (d) 2D plots of the electron density in the two cases are given. We also plot the line out of the on-axis $E_1$ fields for different time points in the boosted frame, as shown in Fig. \ref{fig:sim_lwfa} (e)--(h). As we can see from the comparisons, very good agreement between the results with these two solvers is obtained. 

\begin{table}[hbtp]
\centering
\begin{tabular}{p{8cm}c}
\hline\hline
\textbf{Parameters} & \textbf{Values}\\
\hline
Plasma  &  \\
\quad density $n_p$ & $8.62\times10^{-4} n_0\gamma_b$ \\
\quad length $L$ & $8.0\times10^4 k_0^{-1}/\gamma_b$ \\
Laser &  \\
\quad normalized vector potential $a_0$ & 4.0  \\
\quad focal waist $w_0$  & $153.0k_0^{-1}$  \\
\quad pulse length $\tau$ &  $86.9k_0^{-1}\gamma_b(1+\beta_b)$  \\
\quad polarization & circular \\
Simulation setups & \\
\quad cell size $\Delta x_{1,2,3}$ & 0.1$k_0^{-1}\gamma_b(1+\beta_b)$ \\
\quad time step $\Delta t/\Delta x_1$ & 0.125 \\
\quad number of cells & $2048\times512\times512$ \\
\quad particle shape & quadratic \\
\quad particle per cell & $(2,2,2)$ \\
NCI elimination parameters & \\
\quad Customized solver &  \\
\qquad $[k_1]$ modification $(k_{1l}, k_{1u}, \Delta k_\text{mod,max})$ & $(0.1,0.35,0.01)$ \\
\qquad lowpass filter $(f_l,f_u)$ & (0.3,0.325) \\
\quad Hybrid Yee-FFT solver &  \\
\qquad $[k_1]$ modification $(k_{1l}, k_{1u}, \Delta k_\text{mod,max})$ & $(0.141,0.24,0.007)$ \\
\qquad lowpass filter $(f_l,f_u)$ & (0.3,0.35) \\
\hline\hline
\end{tabular}
\caption{Parameters for a 3D LWFA simulations in the Lorentz boosted frame using the customized high order solver and hybrid Yee-FFT solver. The laser frequency $\omega_0$, wave number $k_0$ and the critical density $n_0=m_e\omega_0^2/(4\pi e^2)$ in the lab frame are used to normalize the simulation parameters. The parameters of $[k_1]$ modification are normalized to $k_{g1}\equiv 2\pi/\Delta x_1$.}
\label{tab:lwfa_pars}
\end{table}

\section{Summary}
\label{sect:summary}
In this paper, we have presented a new customized high-order FDTD solver combined with a current correction (such that Gauss's law remains satisfied) that effectively eliminates the NCI. The current is corrected and filtered by using a local FFT on each parallel partition when using domain decomposition. The customized higher order solver, and the corresponding current correction/filtering that is done locally on each partition, permits the systematic elimination of the Numerical Cerenkov Instability (NCI), while also permitting high parallel scalability in particle-in-cell codes. Using the theoretical framework we developed previously \cite{XuCPC13,YuCPC15} and illustrative PIC simulations, it is found that a high-order FDTD solver has similar NCI properties to that of a fully spectral solver or a hybrid Yee-FFT solver. By reducing the time step, the fastest growing  $(\mu,\nu_1)=(0,\pm1)$ NCI modes and  $(\mu,\nu_1)=(0,0)$ NCI modes can reside very close to the edge of the fundamental Brillouin zone. This enables the use of  a lowpass  filter on the current to effectively eliminate the NCI. For regular high-order FDTD solvers, a highly localized NCI modes [which are part of the $(\mu,\nu_1)=(0,0)$ modes] are seen in analogy to those observed in a spectral or hybrid Yee-FFT solver. These modes reside close to the physical modes in  $\vec k$-space. Elimination of these modes can be achieved by a combination of applying reduced time step and creating a bump in the EM dispersion relation in $k_1$ space. This solver can be readily implemented in 2D/3D Cartesian and quasi-3D geometries contained within the existing framework of OSIRIS without the need to modify the boundary conditions in the transverse directions. We note that the boundary conditions in the $\hat 1$ direction do not need to be changed since we can gradually reducing the order of the solver from 16th to 2nd order in the last 16 cells to match the boundary condition. 

When the finite difference operators are modified, then the charge conserving current deposit must also be appropriately modified. We first deposit the current using the second order accurate  charge conserving current deposit \cite{chargeconservation} in OSIRIS. The current is then Fourier transformed on each local partition, and then corrected, and filtered; it is then transformed back to real space for use in the field solver.  The use of a current deposit that satisfies the continuity equation for the higher order divergence operator is necessary such that Gauss' Law remains satisfied at each time step. We show that making such correction to the current will expand the range of cells over which the current for a particle is increased. Theoretically, a delta function for the current will extend to the entire simulation domain. However, the  current falls below the  double precision roundoff within a finite number of cells. Therefore, the current from a single particle is effectively localized. This permits using FFTs and the current correction and filtering for only the data on each parallel partition if the number of guard cells is properly chosen. 
 
We have shown how the customized solver, together with its NCI elimination scheme, can systematically eliminate the NCI in a single drifting plasma. We have also shown how this scheme can be applied to relativistic shock simulation, with excellent NCI elimination achieved without sacrificing the parallel scalability of an FDTD EM-PIC code for problems with disproportionate number of cells in one direction. We have also shown the usefulness of the proposed high-order solver combined with local FFTs by conducting full 3D LWFA simulations in a Lorentz boosted frame.

\section*{Acknowledgments}
This work was supported by US DOE under grants DE-SC0014260, DE-SC0008316, DE-SC0010064, and by the US National Science Foundation under grant ACI 1339893, OCI 1036224, 1500630, and by NSFC 11425521, 11535006, 11175102, 11375006, and Tsinghua University Initiative Scientific Research Program. Simulations were carried out on the UCLA Hoffman2 and Dawson2 Clusters, and on Edison Cluster of the National Energy Research Scientific Computing Center, and on Blue Waters cluster at National Center for Supercomputing Applications at UIUC.

\begin{appendix}
\section{2D Numerical dispersion for relativistically drifting plasma and NCI analytical expression in customized solver}
\label{sect:app:dispersion}
According to Ref. \cite{XuCPC13,YuCPC15}, the numerical dispersion for the hybrid solver can be expressed as
\begin{align}\label{eq:2dmodesres}
& \left((\omega' - k'_1 v_0)^2- \frac{\omega_p^2}{\gamma^3}(-1)^\mu\frac{S_{j1} S_{E1}\omega'}{[\omega]} \right)\times  \nonumber\\
&\left( [\omega]^2 - [k]_{E1}[k]_{B1}  - [k]_{E2}[k]_{B2}  - \frac{\omega_p^2}{\gamma} (-1)^\mu\frac{S_{j2}(S_{E2}[\omega] - S_{B3} [k]_{E1} v_0)}{\omega' - k'_1 v_0} \right) \nonumber\\
&+ \mathcal{C}=0
\end{align}
where $\mathcal{C}$ is a coupling term in the dispersion relation
\begin{align}\label{eq:coupling}
\mathcal{C}=\frac{\omega_p^2}{\gamma} \frac{(-1)^\mu}{[\omega]}\biggl\{&S_{j1}S_{E1}\omega'[k]_{E2}[k]_{B2}(v^2_0-1)+S_{j2}S_{E2}[k]_{E2}[k]_{B2}(\omega'-k'_1v_0)\nonumber\\
&+S_{j1}[k]_{E2}(S_{E2}[k]_{B1}k_2v_0-S_{B3}k_2v^2_0[\omega])\biggr\}
\end{align}
and for the customized solver discussed in this paper, $[k]_{E1}=[k]_{B1}=[k_1]_{p*}$, where $[k_1]_{p*}$ is defined in Eq. (\ref{eq:k1customize}), and 
\begin{align}
[k]_{E2}=[k]_{B2}=\frac{\sin(k_2\Delta x_2/2)}{\Delta x_2/2}
\end{align}

We can expand $\omega'$ around the beam resonance $\omega' = k'_1v_0$ in Eq. (\ref{eq:2dmodesres}), and write $\omega' = k'_1v_0 + \delta\omega'$, where $\delta\omega'$ is a small term. This leads to a cubic equation for $\delta\omega'$ (see \cite{YuCPC15} for the detailed derivation),
\begin{align}\label{eq:asym1}
A_2\delta\omega'^3+B_2\delta\omega'^2+C_2\delta\omega'+D_2=0
\end{align}
where
\begin{align}
A_2=&2\xi^3_0\xi_1\nonumber\\
B_2=&\xi^2_0\biggl\{\xi^2_0-[k]_{E1}[k]_{B1}-[k]_{E2}[k]_{B2}-\frac{\omega^2_p}{\gamma}(-1)^\mu S_{j2}(S_{E2}\xi_1-\zeta_1S'_{B3}[k]_{E1})\biggr\}\nonumber\\
C_2=&\frac{\omega^2_p}{\gamma}(-1)^\mu \biggl\{ \xi^2_0S_{j2}(\zeta_0S'_{B3}[k]_{E1}-S_{E2}\xi_0)-{\xi_1}S_{j1}[k]_{E2}k_2S_{E2}[k]_{B1}\nonumber\\
& +\xi_0[k]_{E2}(S_{j2}S_{E2}[k]_{B2}-S_{j1}k_2\zeta_1S'_{B3}\xi_0)\biggr\}\nonumber\\
\label{eq:asym2gen}D_2=&\frac{\omega^2_p}{\gamma}(-1)^\mu \xi_0[k]_{E2}k_2S_{j1}\biggl( S_{E2}[k]_{B1}-\zeta_0S'_{B3}\xi_0\biggr)
\end{align}
where
\begin{align}\label{eq:approximation}
\xi_0&=\frac{\sin(\tilde k_1\Delta t/2)}{\Delta t/2}\qquad \xi_1 = \cos(\tilde k_1\Delta t/2)\nonumber\\
\zeta_0&= \cos(\tilde k_1\Delta t/2) \qquad \zeta_1= -\sin(\tilde k_1\Delta t/2)\Delta t/2\nonumber\\
\tilde k_1&=k_1+\nu_1 k_{g1}-\mu\omega_g
\end{align}
We use
\begin{align}
s_{l,i}&=\biggl(\frac{\sin(k_i\Delta x_i/2)}{\Delta x_i/2}\biggr)^{l+1}
\end{align}
as well as use the corresponding interpolation functions for the EM fields used to push the particles 
\begin{align}
S_{E1}&=s_{l,1}s_{l,2}s_{l,3}(-1)^{\nu_1}\qquad S_{E2}=s_{l,1}s_{l,2}s_{l,3} \qquad S_{E3}=s_{l,1}s_{l,2}s_{l,3}\nonumber\\
S_{B1}&=s_{l,1}s_{l,2}s_{l,3}\qquad S_{B2}=s_{l,1}s_{l,2}s_{l,3}(-1)^{\nu_1} \qquad S_{B3}=s_{l,1}s_{l,2}s_{l,3}(-1)^{\nu_1}
\end{align}
when using the momentum conserving field interpolation. The $(-1)^{\nu_1}$ term is due to the half-grid offsets of these quantities in the $\hat 1$ direction. With respect to the current interpolation, \begin{align}
S_{j1} = s_{l-1,1}s_{l,2}s_{l,3}(-1)^{\nu_1}\qquad S_{j2} = s_{l,1}s_{l-1,2}s_{l,3}\qquad S_{j3} = s_{l,1}s_{l,2}s_{l-1,3}
\end{align} 
We note that we use expressions for charge conserving current deposition scheme that are strictly true in the limit of vanishing time step $\Delta t\rightarrow 0$. The coefficients $A_2$ to $D_2$ are real, and completely determined by $k_1$ and $k_2$. By solving Eq. (\ref{eq:asym1}) one can rapidly scan the NCI modes for a particular set of $(\mu,\nu_1)$ for the customized solver.
\end{appendix}

\end{document}